\documentclass[aip,amsmath,amssymb,reprint]{revtex4-1}
\usepackage{graphicx}
\usepackage{dcolumn}
\usepackage{bm}
\usepackage{dsfont}
\usepackage[utf8]{inputenc}
\usepackage[T1]{fontenc}
\usepackage{mathptmx}
\usepackage{etoolbox}
\usepackage{epstopdf}
\usepackage{verbatim}
\usepackage[colorlinks,linkcolor=blue,citecolor=blue,urlcolor=blue]{hyperref}
\usepackage[caption=false,font=scriptsize,labelfont=sf,textfont=sf]{subfig}
\usepackage{multirow}

\makeatletter
\def\@email#1#2{%
 \endgroup
 \patchcmd{\titleblock@produce}
  {\frontmatter@RRAPformat}
  {\frontmatter@RRAPformat{\produce@RRAP{*#1\href{mailto:#2}{#2}}}\frontmatter@RRAPformat}
  {}{}
}%
\makeatother
\begin{document}

\preprint{AIP/123-QED}

\title{High-efficiency and long-distance quantum memory-assisted device-independent quantum secret sharing with single photon sources}
\author{Qi Zhang}
\affiliation{College of Science, Nanjing University of Posts and Telecommunications, Nanjing, Jiangsu 210023, China}
\author{Jia-Wei Ying}
\affiliation{College of Electronic and Optical Engineering and College of Flexible Electronics (Future Technology), Nanjing University of Posts and Telecommunications, Nanjing, Jiangsu 210023, China}
\author{Shi-Pu Gu}
\affiliation{College of Electronic and Optical Engineering and College of Flexible Electronics (Future Technology), Nanjing University of Posts and Telecommunications, Nanjing, Jiangsu 210023, China}
\author{Xing-Fu Wang}
\affiliation{College of Science, Nanjing University of Posts and Telecommunications, Nanjing, Jiangsu 210023, China}
\author{Ming-Ming Du}
\affiliation{College of Electronic and Optical Engineering and College of Flexible Electronics (Future Technology), Nanjing University of Posts and Telecommunications, Nanjing, Jiangsu 210023, China}
\author {Wei Zhong}
\affiliation{Institute of Quantum Information and Technology, Nanjing University of Posts and Telecommunications, Nanjing, Jiangsu 210003, China}
\author{Lan Zhou}
\altaffiliation{Authors to whom correspondence should be addressed: zhoul@njupt.edu.cn }
\affiliation{College of Science, Nanjing University of Posts and Telecommunications, Nanjing, Jiangsu 210023, China}
\author{Yu-Bo Sheng}
\affiliation{College of Electronic and Optical Engineering and College of Flexible Electronics (Future Technology), Nanjing University of Posts and Telecommunications, Nanjing, Jiangsu 210023, China}

\date{\today}

\begin{abstract}
Quantum secret sharing (QSS) plays a critical role in building the distributed quantum networks. Device-independent (DI) QSS provides the highest security level for QSS. However, the photon transmission loss and extremely low multipartite entanglement generation rate largely limit DI QSS's secure photon transmission distance (less than 1 km) and practical key generation efficiency. To address the above drawbacks, we propose the quantum memory-assisted (QMA) DI QSS protocol based on single photon sources (SPSs). The single photons from the SPSs are used to construct long-distance multipartite entanglement channels with the help of the heralded architecture. The heralded architecture enables our protocol to have an infinite secure photon transmission distance in theory. The QMA technology can not only increase the multi-photon synchronization efficiency, but also optimize the photon transmittance to maximize the construction efficiency of the multipartite entanglement channels. Our protocol achieves the practical key generation efficiency seven orders of magnitude higher than that of the existing DI QSS protocols based on cascaded spontaneous parametric down-conversion sources and six orders of magnitude higher than that of the DI QSS based on SPSs without QMA. Our protocol has modular characteristics and is feasible under the current experimental technical conditions. Combining with the advanced random key generation basis strategy, the requirement on experimental devices can be effectively reduced. Our protocol is expected to promote the development of long-distance and high-efficiency DI quantum network in the future.
\end{abstract}

\maketitle

As an important branch of quantum communication, quantum secret sharing (QSS) ensures that multiple players cooperate to reconstruct the dealer's secret \cite{QSS1}. QSS holds significant applications in future quantum networks. Since the first QSS protocol in 1999 \cite{QSS1}, QSS has flourished with important theoretical advancements and experimental demonstrations \cite{QSS2,QSS3,MDIQSS1,MDIQSS2,QSSe1,QSSe2,QSSe3,QSSe4}. The unconditional security of general QSS protocols \cite{QSS1,QSS2,QSS3} depends on the assumption about perfect experimental devices. However, the practical imperfect devices may be susceptible to side-channel attacks. In recent years, device-independent (DI) QSS \cite{DIQSS1,DIQSS2,DIQSS3,DIQSS4} has emerged for enhancing the QSS's robustness against device vulnerabilities.

DI-type protocols originated from DI quantum key distribution (QKD) \cite{DIQKD1,DIQKD2,DIQKD3}. DI protocols rely on the joint probability distributions to evaluate the quantum nonlocal correlations between the shared particles. This allows the experimental devices to be treated as black boxes, with no need to trust their internal workings. In this way, the DI protocols can provide the highest security level for quantum communication. Over the past decade, DI QKD has achieved a series of important theoretical \cite{DIQKD4,DIQKD5,DIQKD6,DIQKD7,DIQKD8,DIQKD9,DIQKD10,DIQKD11,DIQKD12} and experimental advancements \cite{DIQKDe1,DIQKDe2,DIQKDe3}. The research on DI QSS originated in 2019 \cite{DIQSS1,DIQSS2}. Since 2024, DI QSS protocols in practical communication scenarios have been proposed \cite{DIQSS3,DIQSS4}. Meanwhile, the active improvement strategies have been introduced in DI QSS to reduce the experimental difficulty. Existing DI QSS protocols \cite{DIQSS3,DIQSS4} require the central source to generate the multipartite entangled photons, such as the Greenberger-Horne-Zeilinger (GHZ) state, through the cascaded spontaneous parametric down-conversion (SPDC) processes \cite{GHZ1} and distribute the entangled photons to multiple users. The cascaded SPDC sources generate the GHZ state with a quite low rate ($10^{-10}$-$10^{-6}$) \cite{GHZ2}. Moreover, entangled photon pairs are highly susceptible to noise during long-distance entanglement distribution. Photon transmission loss leads to a significant degradation of the quantum nonlocal correlations. These two obstacles severely reduce DI QSS's practical key generation efficiency and secure communication distance. Even with a series of active improvement strategies, DI QSS's maximal secure communication distance is only about 1.41 km \cite{DIQSS3,DIQSS4}.

Improving the entanglement channel's construction efficiency and noise robustness is essential for developing high-efficiency and long-distance DI QSS. In 2014, the construction protocol for the long-distance two-photon entanglement channel with realistic single-photon sources (SPSs) and heralded architecture was proposed \cite{Heralded0}. The practical SPS can generate the single photon in an almost on-demand way \cite{source0,source1,source2,source3}, and the heralded architecture can eliminate the influence of photon transmission loss on entanglement channel's quality. $Ko{\l}ody\'{n}ski$ \emph{et al.} introduced that method into DI QKD to propose the DI QKD based on SPSs, which is called SPS DI QKD \cite{Heralded1,Heralded2}. The critical step is that each party couples two single photons with orthogonal polarization into one spatial mode and transmits only one photon to the distant measurement station for the heralded Bell state measurement (BSM). However, for constructing the high-purity entanglement channel, the photon transmittance has to approach zero ($\sim 10^{-3}$) \cite{Heralded1,Heralded2}. Meanwhile, the photon transmission loss largely reduces the two-photon synchronization efficiency for the BSM. Both these factors lead to quite low entanglement channel construction efficiency, and thus severely reduce SPS DI QKD's practical key generation efficiency.

Inspired by the SPS DI QKD, we propose the high-efficiency and long-distance quantum-memory-assisted (QMA) DI QSS based on SPSs, which is called the QMA SPS DI QSS protocol. We introduce the heralded method in ref. \cite{Heralded0} to construct the high-quality multipartite entanglement channels from single photons. The influence of photon transmission loss on channel quality can be eliminated by the heralded architecture. The QMA technology has been widely used in measurement-device-independent (MDI) QKD and MDI QSS systems to improve the synchronization efficiency at the center measurement station \cite{MDIQSS2,QMA1,QMA2,QMA3,QMA4,QMA5}. In our protocol, the adoption of the QMA technology can not only increase the photon synchronization efficiency, but also optimize the photon transmittance to maximize the construction efficiency of the multipartite entanglement channels. The QMA SPS DI QSS protocol has an infinite secure photon transmission distance in theory. Moreover, it achieves the practical key generation efficiency seven orders of magnitude higher than that of the DI QSS protocol with cascaded SPDC sources (SPDC DI QSS protocol) \cite{DIQSS3,DIQSS4}, and six orders of magnitude higher than that of the SPS DI QSS without the QMA. Our protocol is also compatible with active improvement strategies, which can relax the performance requirements for local devices. Our protocol's setup is modular, making it well-suited for extending to multi-user scenarios, and each module is feasible under current experimental conditions. Our QMA SPS DI QSS protocol provides a possible way to realize the high-efficiency and long-distance DI quantum network in the future.

\begin{figure*}
\includegraphics[scale=0.39]{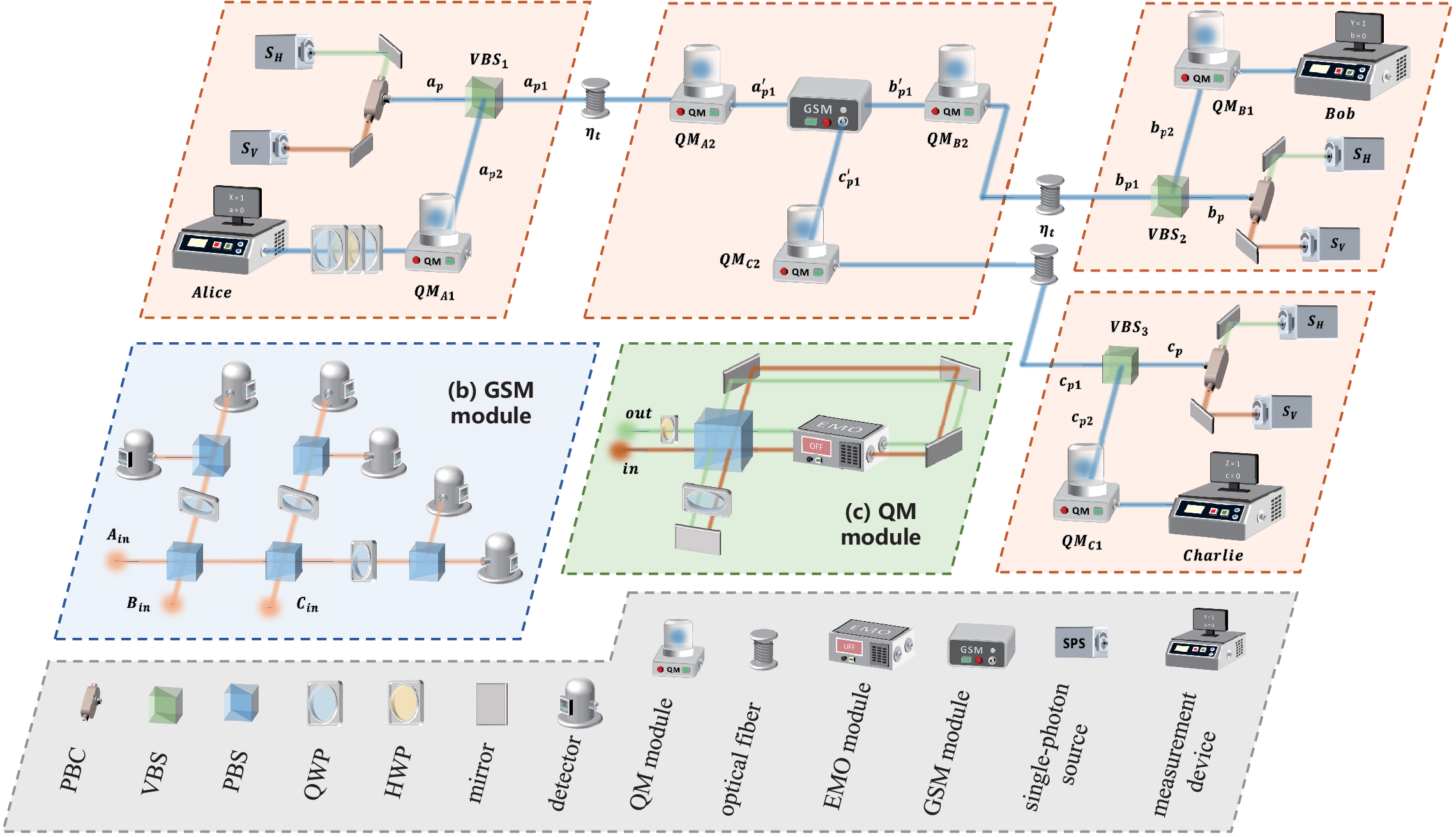}
\caption{(a) Schematic diagram of the QMA SPS DI QSS protocol. The single-photon sources $S_H$ and $S_V$ generate single photons in the horizontal polarization $|H\rangle$ and vertical polarization $|V\rangle$, respectively. The half-wave plate (HWP) realizes $|H\rangle\leftrightarrow|V\rangle$. The quarter-wave plate (QWP) realizes $|H\rangle\rightarrow\frac{1}{\sqrt{2}}(|H\rangle+|V\rangle)$ and $|V\rangle\rightarrow\frac{1}{\sqrt{2}}(|H\rangle-|V\rangle)$. (b) The structure of the three-photon GSM module \cite{GSM1}. The polarization beam splitter (PBS) can totally transmit $|H\rangle$ polarized photon and reflect $|V\rangle$ polarized photon. The single-photon detectors are used to detect the photons in different output modes. (c) Schematic diagram of the all-optical storage loop QM module \cite{QM1}. We can control the storage and output of single photon by controlling the "on-off" of the electro-optic modulator (EOM).}
\label{fig1}
\end{figure*}

Here, we consider the three-partite QMA SPS DI QSS protocol. The three users include the dealer Alice and two players, Bob and Charlie. The schematic diagram of the protocol and the structures of the GHZ state measurement (GSM) and QM modules are shown in Fig.~\ref{fig1}. The protocol includes six steps as follows.

\textbf{Step 1}. Each user generates two single photons in $|H\rangle$ and $|V\rangle$ from SPSs named $S_H$ and $S_V$, respectively. The users pass the generated photons through the polarization beam combiners (PBCs) into the spatial modes $a_p$, $b_p$, and $c_p$, respectively. The PBC can erase the original path information of the two photons with different polarization states.

\textbf{Step 2}. Each user passes the two converging photons through a variable beam splitter (VBS) with the transmittance of $T$. The transmitted photons pass through the lossy channels with the transmission efficiency of $\eta_t$ and are stored in the $QM_{A2}$, $QM_{B2}$, and $QM_{C2}$ at the center measurement station, respectively. The reflected photons are directed into the $QM_{A1}$, $QM_{B1}$, and $QM_{C1}$, respectively. Each QM continuously monitors the arrival of a single photon and replaces the stored photon with a newer arriving one.

\textbf{Step 3}. The event that each of the six QMs is loaded with a single photon heralds the successful separation of the two photons in each user's location and the successful arrival of all the transmitted photons. For each user, the success probability can be calculated as $P_s=2\eta_t T(1-T)$. Then, the measurement party extracts the single photons from $QM_{A2}$, $QM_{B2}$, and $QM_{C2}$ to the GSM module (Fig.~\ref{fig1} (b)) from the spatial modes $a_{p1}'$, $b_{p1}'$, and $c_{p1}'$, respectively. Only when the GSM is successful, the photons stored in $QM_{A1}$, $QM_{B1}$, and $QM_{C1}$ can successfully establish the entanglement correlations, which is called the successful entanglement event. The details are shown in Appendix I.

\textbf{Step 4}. After the successful construction of three-partite entanglement channels, the three users read out the stored photons from $QM_{A1}$, $QM_{B1}$, and $QM_{C1}$. Then, each user randomly selects measurement basis to measure the photon. Alice and Charlie have two basis choices $A_i$ and $C_k$ ($i,k\in\{1,2\}$), where $A_1=C_1=\sigma_x$, and $A_2=-C_2=\sigma_y$. Bob has three basis choices $B_j$ ($j\in\{1,2,3\}$), where $B_1=\sigma_x$, $B_2=\frac{\sigma_x-\sigma_y}{\sqrt{2}}$, and $B_3=\frac{\sigma_x+\sigma_y}{\sqrt{2}}$. We denote that each of the measurement bases has two possible outputs $a_i,b_j,c_k\in\{-1,+1\}$. After all photons are measured, Alice, Bob, and Charlie announce their basis choices. Suppose that Bob selects $j\in\{2,3\}$ with the probability of $P_c$.

\emph{Discarded rounds}: When the measurement basis combination is $\{A_1B_1C_2\}$, $\{A_2B_1C_2\}$, or $\{A_2B_1C_1\}$, three users have to discard their corresponding measurement results.

\emph{Key generation rounds}: When $i=k=j=1$, the three users' measurement results are highly correlated. The three users preserve their measurement results as the key bits. We label the measurement result $+1$ as the key bit 0, $-1$ as the key bit 1. The coding rule is $k_A=k_B\oplus k_C$, where $k_A$, $k_B$, and $k_C$ denote the key bits of Alice, Bob, and Charlie, respectively.

\emph{Security test rounds}: When $i,k\in\{1,2\}$ and $j\in\{2,3\}$, all the users announce their measurement results to perform the Svetlichny test. The violation of Svetlichny inequality (The Svetlichny polynomial satisfies $S_{ABC}>4$) can determine the genuine three-photon quantum nonlocality \cite{Svetlichny,DIQSS3,DIQSS4}, and the protocol goes to the next step. $S_{ABC}>4$ is equivalent to Alice's and Bob's measurement results violating the CHSH inequality (the CHSH polynomial satisfies $S>2$) \cite{Bancal}. If $S_{ABC}\leq 4$ (equivalent to $S\leq2$), the security test is not passed. In this case, all the key bits have to be discarded.

\emph{\textbf{Step 5}}. The three users repeat the above steps until they obtain sufficient key bits. Then, they perform the error correction and private amplification to distill the secure key bits.

\emph{\textbf{Step 6}}. Charlie publishes his subkey $k_C$, and Bob can reconstruct the key $k_A$ delivered by Alice combining $k_C$ with his own subkey $k_B$.

Then, we estimate the practical key generation efficiency $E_c$ of our QMA SPS DI QSS protocol in the noisy environment. Similar to all the DI-type protocols \cite{DIQKD1,DIQKD2,DIQKD3,DIQSS3,DIQSS4,DIQSDC1}, the QMA SPS DI QSS protocol relies on only two basic assumptions: the correctness of quantum physics and the security of the users' physical locations. The legitimacy and honesty of the three users in the key generation stage are also essential prerequisites for security. In the security analysis, we do not impose any limitations on the ability of the eavesdropper (Eve), who can even take full control of the SPSs and the user's measurement devices. The violation of Svetlichny inequality guarantees the randomness of the device's output results, and thus ensures the key's uncertainty to Eve.

We adopt the all-optical polarization-insensitive storage loop QM \cite{QM1} with the structure of Fig.~\ref{fig1} (c). Here, we configure that each QM can store a single photon for at most $N$ photon pulse intervals. We analyze the efficiency $E_m$ that each of the six QMs is loaded by a single photon per unit time.

We assume that the three users achieve successful two-photon separation events at the $(l+1)$th, $(m+1)$th, $(n+1)$th pulse intervals, respectively, where $l,m\leq n$ and the actual storage pulse interval $n\leq N$. In this way, the probability that all the six QMs are fully loaded with photons at the ($n+1$)th photon pulse interval can be calculated as
\begin{eqnarray}\label{Pn+1}
P(n+1) = P_s^3 (1-P_s)^n \left[ \frac{(\eta_M^2 (1-P_s))^n - 1}{1 - \frac{1}{\eta_M^2 (1-P_s)}} + 1 \right]^2,
\end{eqnarray}
where $\eta_M$ denotes the storage efficiency of the QM. For all $n\leq N$ cases, the total fully loaded probability is
\begin{eqnarray}\label{PtN+1}
P_t(N+1)=\sum_{n=0}^{N}P(n+1).
\end{eqnarray}

Since the number of consumed photon pulses is proportional to $n$, we need to consider the average photon pulse consumption per unit time, denoted as $P_w$, which represents the statistical average over all $n$. In the first case, we consider $n<N$. In this case, the three users complete one QM fully loaded event in advance, with each SPS transmitting $n+1$ photon pulses. When $n<N$, the photon pulse interval consumption is
\begin{eqnarray}\label{PwN}
P_w(\Sigma N)=\sum_{n=0}^{N-1}(n+1)P(n+1).
\end{eqnarray}
In the second case, we consider $n=N$. Here, regardless of whether all QMs are successfully loaded with photons, each SPS must emit $N+1$ photon pulses. In this case, the photon pulse interval consumption is
\begin{eqnarray}\label{PwN+1}
P_w(N+1)=(N+1)(1-P_t(N)).
\end{eqnarray}
Finally, we obtain the total photon pulse interval consumption as $P_w=P_w(\Sigma N)+P_w (N+1)$. Accordingly, the fully loaded efficiency $E_m$ can be calculated as
\begin{eqnarray}\label{Em}
E_m=\frac{P_t(N+1)}{P_w(\Sigma N)+P_w(N+1)}.
\end{eqnarray}
The details are shown in Appendix II A.

\begin{figure}
\includegraphics[scale=0.34]{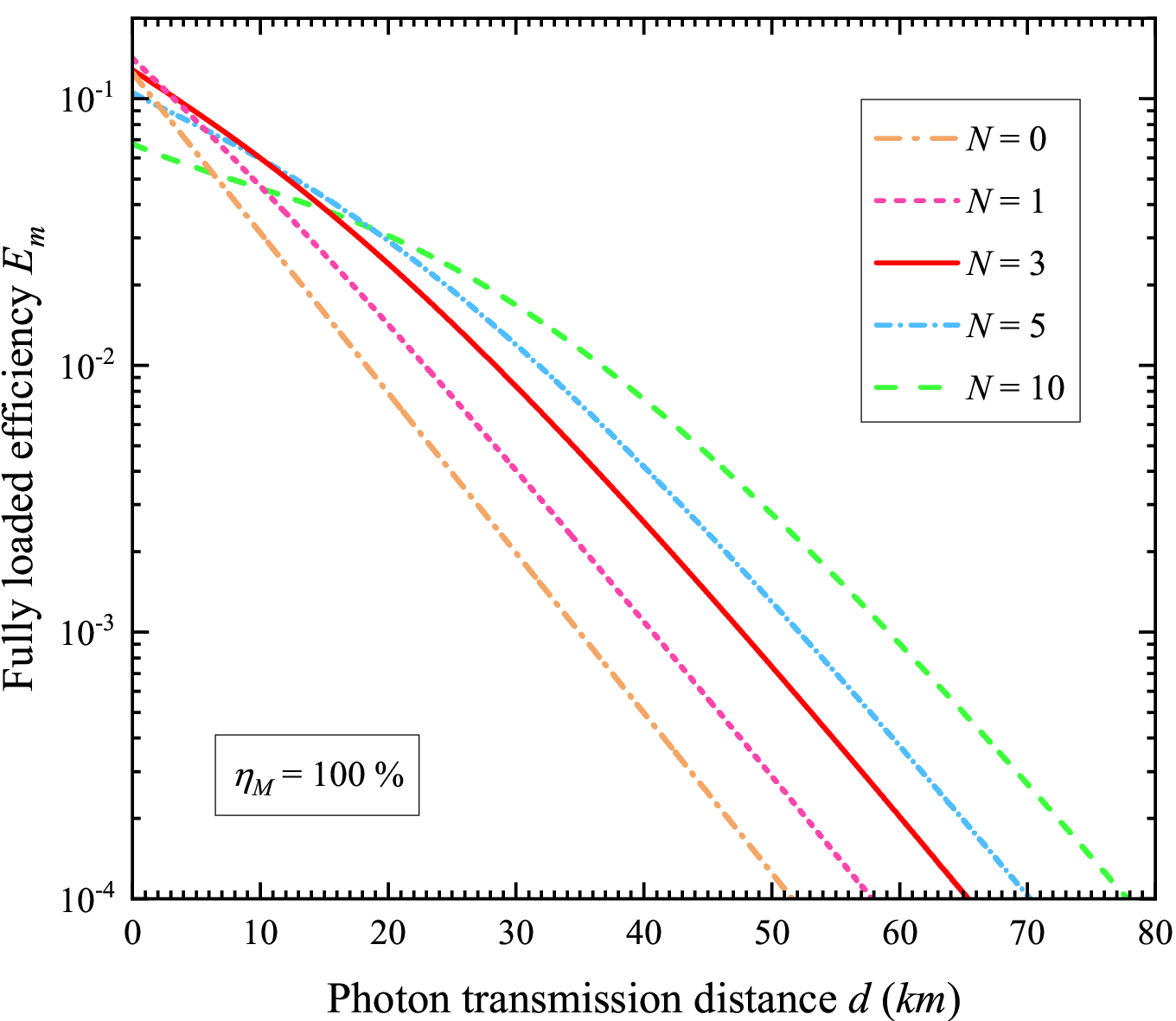}
\caption{The fully loaded efficiency $E_m$ as a function of photon transmission distance $d$, where the QM storage efficiency is fixed at $\eta_M=100\%$, and the maximum storage pulse interval $N$ is set as $N=0,1,3,5,10$.}
\label{fig2}
\end{figure}

It can be calculated that $E_m$ reaches the maximum when $T=0.5$ (see Appendix II A). In Fig.~\ref{fig2}, we provide $E_{m}$ as a function of the photon transmission distance $d$ with $T=0.5$ and $\eta_M=100\%$. The maximum storage pulse interval is set as $N=0, 1, 3, 5, 10$. Here, the case $N=0$ represents that the QMs only herald successful separation events without storing the photons. It can be found that as $N$ increases, $E_m$ improves in the long-distance scenario, but $E_m$ reduces in the  short-distance scenario. In the short-distance scenario (the photon transmission loss is low), the high value of $N$ leads to weak synchronization efficiency growth but large time consumption, so that $E_m$ reduces with the growth of $N$. However, in the long-distance scenario (the photon transmission loss is high), with the growth of $N$, the advantage of synchronization efficiency growth exceeds the disadvantage of time consumption, which leads to the growth of $E_m$.

In our QMA SPS DI QSS protocol, the practical key generation efficiency $E_c$ (The practical key generation rate per second) can be calculated by
\begin{eqnarray}\label{Ec1}
E_c=(1-P_c)P_{GHZ}R_{rep}E_m R_\infty,
\end{eqnarray}
where $R_{rep}$ represents the repetition frequency of the on-demand SPS, $P_{GHZ}$ is the success probability of the GSM module. In the asymptotic limit of a large number of rounds, the total secure key rate $R_\infty$ is defined as the ratio of the extractable key length to the number of key generation rounds.

Here, we suppose that Alice (Charlie) chooses $A_1$ and $A_2$ ($C_1$ and $C_2$) with probabilities $p$ and $\bar{p}=1-p$, respectively. Based on Ref. \cite{DIQSS3}, $R_\infty$ of the our protocol under the collective attack is lower bounded by
\begin{eqnarray}\label{rinfty2}
R_\infty\geq p^2\left[g\left(\sqrt{S^2/4-1}\right)-h(\delta)\right],
\end{eqnarray}
where $g(x)=1-h(\frac{1}{2}+\frac{1}{2}x)$, and the binary Shannon entropy $h(x)=-x\log_{2}{x}-(1-x)\log_{2}{(1-x)}$. $\delta$ represents the total quantum bit error rate (QBER). The details are shown in Appendix II B.

In this way, we obtain the lower bound of $E_c$ as
\begin{eqnarray}\label{EC2}
E_c \geq \frac{1}{8}R_{rep}E_mp^2\left[g\left(\sqrt{S^2/4-1}\right)-h(\delta)\right].
\end{eqnarray}

During the practical implementation, although the influence from photon transmission loss on the entanglement channel can be eliminated by the heralded architecture, the local devices may cause the photon local loss with probability of $\bar{\eta_l}=1-\eta_l$, and the channel noise may degrade the target GHZ state into eight possible GHZ states with equal probability $\frac{1-F}{8}$. We can estimate $S=2\sqrt{2}F\eta_l^3$ and $\delta=1-\frac{1}{2}\eta_l^3-\frac{1}{2}\eta_l^3F$. Here, we set $P_c=50\%$ and $R_{rep}=10$ MHz \cite{source1}. We use the standard linear optical GSM module as shown in Fig.~\ref{fig1} (b) \cite{GSM1}, which can only identify two GHZ states ($|GHZ_1^\pm\rangle$) with the success probability is $P_{GHZ}=1/4$.

It is natural that $R_\infty$ and $E_c$ decrease with the reduction of fidelity $F$ \cite{DIQSS3,DIQSS4}. To obtain a positive $R_\infty$, the critical value of $F$ is about $81.54\%$. Within the critical scale of $F$, the photon transmission efficiency $\eta_t=10^{-\alpha d/10}$ only influences the value of $E_m$, and is independent of $R_\infty$ and $E_c$, where $d$ denotes the photon transmission distance and $\alpha=0.2$ dB/km for standard optical fiber. In this way, our QMA SPS DI QSS protocol can maintain the positive practical key generation efficiency even at infinite distance.

\begin{figure}
\includegraphics[scale=0.34]{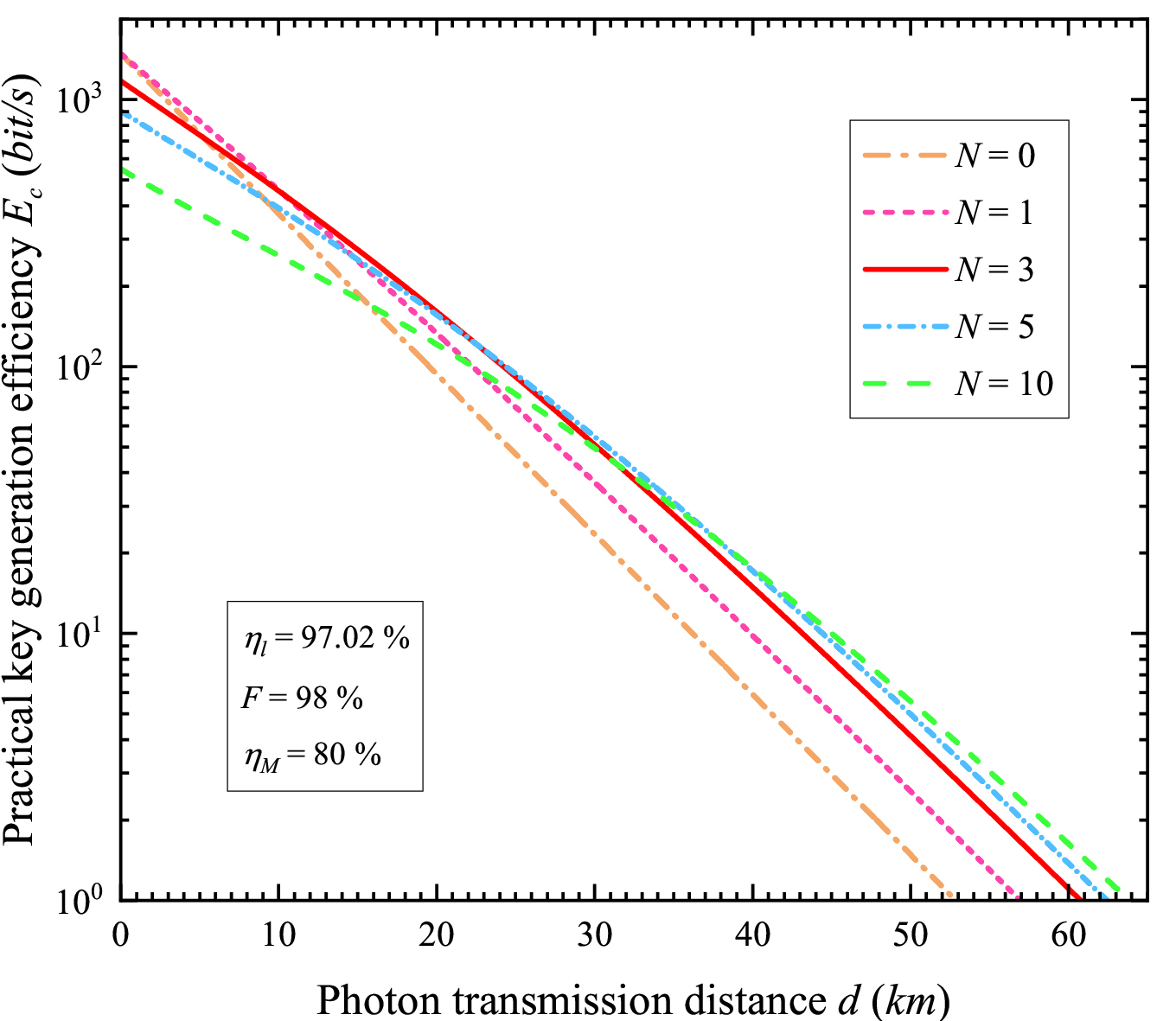}
\caption{The practical key generation efficiency $E_c$ as a function of the photon transmission distance $d$. Here, we fix the fidelity $F=98\%$, the local efficiency $\eta_l=97.02\%$, the storage efficiency $\eta_M=80\%$, and the maximum storage pulse interval $N=0,1,3,5,10$.}
\label{fig3}
\end{figure}

Figure \ref{fig3} demonstrates the lower bound of $E_c$ as a function of the photon transmission distance $d$. Similar to Fig.~\ref{fig2}, in the long-distance scenario, $E_c$ improves with the growth of $N$, but the improvement becomes negligible once $N\geq 5$. In the short-distance scenario, due to the relatively high transmission efficiency $\eta_t$, increasing $N$ reduces $E_m$ and thus leads $E_c$ to decrease. Considering both long-distance and short-distance scenarios comprehensively, $N=3$ (red line) is an optimized value. When $N=3$, three users can generate secure keys  of 1 bit/s at the photon transmission distance of about 60.77 km, extending the secure photon transmission distance by 7.90 km corresponding to the case $N=0$.

The local efficiency $\eta_l=\eta_c \eta_d$, where $\eta_c$ is the coupling efficiency between photon and optical fiber, and $\eta_d$ is the detection efficiency of the photon detector. For obtaining the positive $R_\infty$, the threshold of $\eta_l$ is extremely high. For example, with the parameters in Fig.~\ref{fig3}, the threshold of $\eta_l$ is as high as 96.32\% \cite{DIQSS3}. For reducing the requirement for experimental devices, we can combine the QMA SPS DI QSS protocol with the  active improvement strategies.

We adopt the advanced random key generation basis strategy \cite{DIQSS4} in the QMA SPS DI QSS protocol, which combines the noise preprocessing, postselection, and random key generation basis strategies. The advanced random key generation basis strategy can effectively increase Eve's total uncertainty about the key generation basis and measurement results, thus increasing the total secure key rate \cite{DIQSS4}. The practical key generation efficiency $E_c^{ar}$ of QMA SPS DI QSS protocol with advanced random key generation basis strategy is given by
\begin{eqnarray}\label{ECar}
E_c^{ar} \geq \frac{1}{8}R_{rep}E_m\left(p^2+\bar{p}^2\right) \bigg[g\left(\tilde{E}_\lambda\left(S\right),q\right)-h\left(\delta_{ar}\right)\bigg],
\end{eqnarray}
where $\delta_{ar}$ is the total QBER with the advanced random key generation basis strategy, $\tilde{E}_\lambda(S)$ is the optimal solution corresponding to Eve's total uncertainty about Alice's key, and $g(x,q)$ is the noise preprocessing entropy function \cite{DIQSS4}. When the noise preprocessing level $q\rightarrow50\%$, the local efficiency threshold $\eta_l$ decreases from 96.32\% to 93.41\%, demonstrating an effectively improvement in local loss robustness. The details are shown in Appendix III C.

In Fig.~\ref{fig4}, we compare the practical key generation efficiency of various DI QSS protocols. For the SPDC DI QSS protocol \cite{DIQSS3}, the maximal secure secure photon distance is only about 0.047 km. The secure photon distance corresponding to $E_c=10^{-4}$ bit/s is about 0.037 km (orange line). In the other three DI QSS protocols, the heralded architectures extend their maximal secure photon transmission distance to infinity in theory. In the SPS DI QSS protocol without QMA (blue line), due to the extremely low photon transmittance ($T\approx 10^{-3}$) requirement, its improvement in $E_c$ comparing to the SPDC DI QSS protocol is quite limited (only about 5.34 times). The secure secure photon distance corresponding to $E_c=10^{-4}$ bit/s is about 23.46 km. On the contrary, benefit to the QMA technology, $E_c$ of the QMA SPS DI QSS protocol with $N=3$ (green line) achieves six orders of magnitude higher than that of the SPS DI QSS protocol without QMA, and the secure photon distance corresponding to $E_c=10^{-4}$ bit/s extends to 128.19 km. Moreover, combing with the advanced random key generation basis strategy, $E_c$ can be further increased by about 8 times, and the secure photon distance at $E_c=10^{-4}$ bit/s can be further extended to 152.69 km.

\begin{figure}
\includegraphics[scale=0.34]{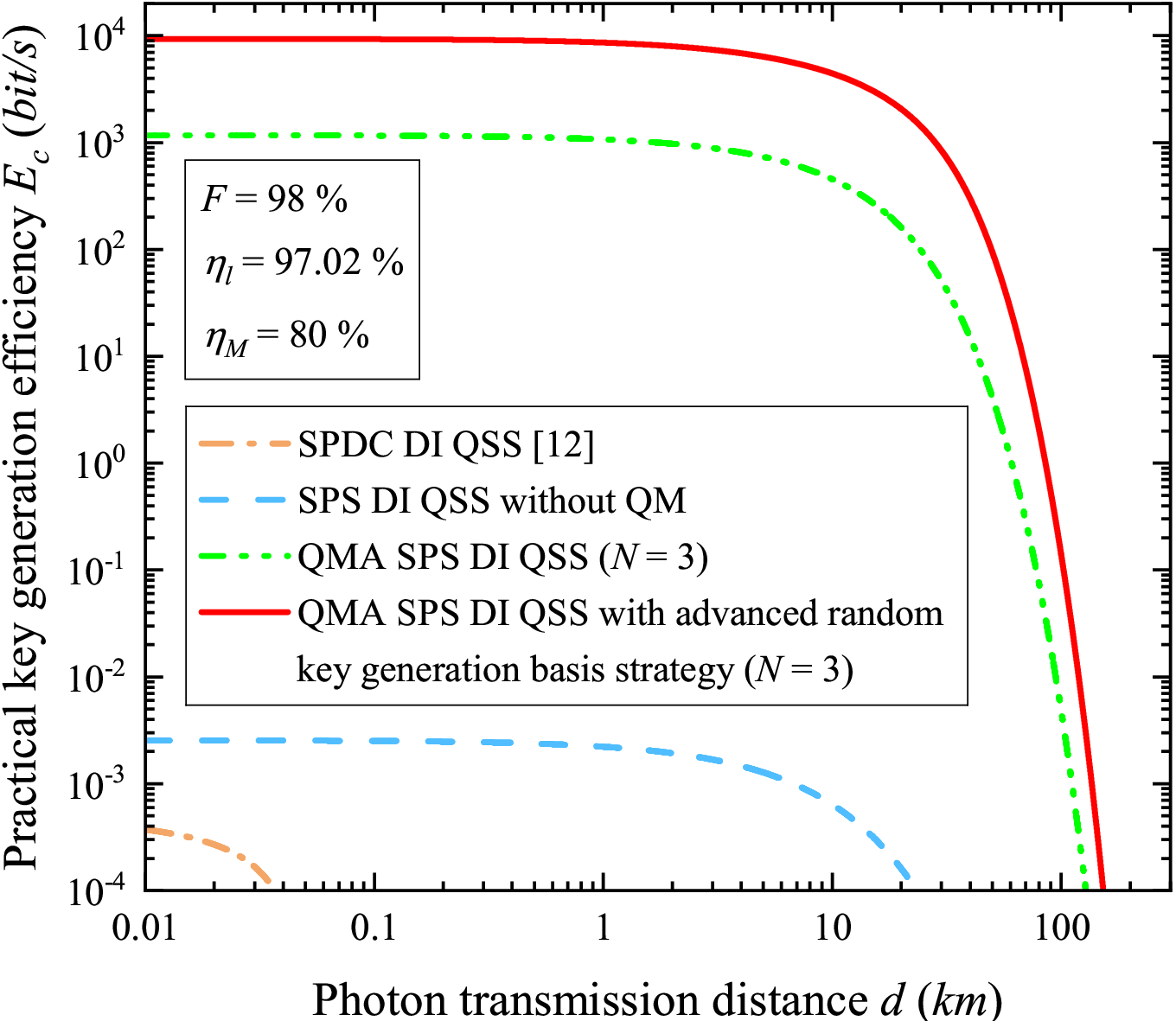}
\caption{The practical key generation efficiency $E_c$ of various DI QSS protocols as a function of photon transmission distance $d$. Here, the fidelity is set to $F=98\%$ and the local efficiency to $\eta_l=97.02\%$. For the SPDC DI QSS protocol \cite{DIQSS3}, the probability of generating a three-photon GHZ state is $10^{-8}$. For QMA SPS DI QSS, the storage efficiency is set to $\eta_M=80\%$.}
\label{fig4}
\end{figure}

Our QMA SPS DI QSS protocol can be divided into the photon generation and separation module, the user's measurement modules, the linear-optical GSM module, and the QM modules. It is natural to extend the three-user protocol to the general M-user protocol ($M>3$) by increasing the number of the photon generation and separation modules, measurement modules and QM modules. The above key modules of the QMA SPS DI QSS protocol are achievable with current experimental conditions. In detail, practical single-photon sources already achieve high-efficiency, on-demand single-photon emission \cite{source2,source3}. Single-photon sources have performed well in ensuring photon purity and indistinguishability with the probability of above 99\% with existing technologies \cite{source2,source3}. Recently, telecommunication-wavelength single-photon sources based on InAs/GaAs quantum dots have achieved count rates up to 10 MHz \cite{source1}. The construction of multipartite entanglement channels is similar to the multiparty quantum repeater with the help of the GSM and QMs. For reducing the experimental cost and enhancing our protocol's practical feasibility, in the QM module, we adopt the all-optical polarization-insensitive storage loop QM. This all-optical storage loop QM has the storage efficiency of 91\% and the lifetime of 131 ns for the photons with the center wavelength of 1550 nm and the bandwidth of 0.52 THz \cite{QM1}. The lifetime of 131 ns is sufficient for a proof-of-principle demonstration of the QMA SPS DI QSS. In addition, Ref. \cite{coupling} reduced the interconnection loss between a nested antiresonant nodeless type hollow-core fiber and a standard single-mode fiber, resulting in the coupling efficiency $\eta_c$ of 96.61\%. The superconducting nanowire single-photon detector with detection efficiency $\eta_d$ of 98\% in the 1550 nm band was reported \cite{SNSPD}. These achievements lead to $\eta_l=\eta_c\eta_d\approx94.7\%$, which is lower than the threshold of $\eta_l$ (93.41\%) of our QMA SPS DI QSS protocol with advanced key generation basis strategy. Therefore, it is possible to realize the experimental demonstration of our QMA SPS DI QSS protocol under the current experimental conditions.

There are some methods to improve the performance of our QMA SPS DI QSS protocol. Firstly, the QM's lifetime of 131 ns would limit the photon transmission distance. For realizing the long-distance QMA SPS DI QSS, we can adopt the solid-state QMs \cite{QM2,QM3,QM4,QM5,QM6,QM7}, which have much longer storage life.
 For example, the atomic QMs in Refs. \cite{QM4,QM5,QM6} achieve the storage lifetime of approximately microseconds.
Recently, the coherent electromechanical interface based on cubic silicon-carbide membrane crystal enables the storage of photons exceeding an hour \cite{QM7}. The long storage time of QM enables our protocol to realize long-distance QMA SPS DI QSS. The linear-optical GSM has the success probability of 1/4, which limits the value of $E_c$. We can adopt the complete GSM module based on hyperentanglement-assisted \cite{GSM2} or quantum non-destructive measurement \cite{GSM3} to achieve the identification of eight GHZ states, which are expected to improve $E_c$ by four times. In the key generation stage of Step 4, pre-shared keys can also be employed to further improve $E_c$ \cite{DIQKD11}.

In conclusion, we propose the high-efficiency and long-distance QMA SPS DI QSS protocol. The SPS can generate single photons in an almost on-demand way, which are used to construct the long-distance entanglement channel. The heralded architecture based on QM and GSM can optimize the construction efficiency of the multipartite entanglement channels. Our QMA SPS DI QSS protocol has the following advantages. First, the heralded architecture can eliminate the influence of photon transmission loss on the multipartite entanglement channel quality, which enables our protocol to have infinite secure photon transmission distance in theory. Second, the QMA technology is used to effectively enhance the synchronization efficiency of the central GSM module and optimize the photon transmittance at each user's location. These optimizations lead our QMA SPS DI QSS protocol to achieve the practical key generation efficiency seven orders of magnitude higher than that of the SPDC DI QSS protocol \cite{DIQSS3,DIQSS4} and six orders of magnitude higher than that of the SPS DI QSS without QMA. Third, our protocol has modular characteristics, being suitable for extending to the multi-user QMA SPS DI QSS, and each module is feasible under the current experimental technical conditions. Finally, we combine our protocol with the advanced random key generation basis strategy to further reduce the requirement on experimental devices. It is expected to promote the development of high-efficiency and long-distance DI quantum networks in the future.

\textbf{Supplementary Material}\\
Long-distance channel construction, practical key generation efficiency, and relaxing the performance requirement for local devices.

\textbf{Acknowledgment}
This work was supported by the National Natural Science Foundation of China under Grants No. 12175106, 92365110, and 12574393, the Postgraduate Research \& Practice Innovation Program of Jiangsu Province under Grant No.KYCX25-1245.

\textbf{Author Declarations}\\
\textbf{Conflict of Interest} The authors have no conflicts to disclose.

\textbf{Data Availability}
The data that support the findings of this study are available within the article (and its supplementary material).

\end{document}


\title{Supplementary Material\\High-efficiency and long-distance quantum memory-assisted device-independent quantum secret sharing with single photon sources}
	
\author{Qi Zhang$^{1}$, Jia-Wei Ying$^{2}$, Shi-Pu Gu$^{2}$, Xing-Fu Wang$^{1}$, Ming-Ming Du$^{2}$, Wei Zhong$^{3}$, Lan Zhou$^{1}$\footnote{Email address: zhoul@njupt.edu.cn}, Yu-Bo Sheng$^{2}$ }
\address{
$^1$College of Science, Nanjing University of Posts and Telecommunications, Nanjing, Jiangsu 210023, China\\
$^2$College of Electronic and Optical Engineering and College of Flexible Electronics (Future Technology), Nanjing University of Posts and Telecommunications, Nanjing, Jiangsu 210023, China\\
$^3$Institute of Quantum Information and Technology, Nanjing University of Posts and Telecommunications, Nanjing, Jiangsu 210003, China}
\date{\today}
\maketitle

\section{Construction of the long-distance entanglement channel}
\emph{\textbf{Step 1}}. Each user generates two single photons in $|H\rangle$ and $|V\rangle$ from SPSs named $S_H$ and $S_V$, respectively. The users pass the generated photons through the PBCs into the spatial modes $a_p$, $b_p$, and $c_p$, respectively. The PBC can erase the original path information of the two photons with different polarization states.

\emph{\textbf{Step 2}}. Each user passes the two converging photons through a VBS with the transmittance of $T$, the quantum state of each two-photon system evolves to
\begin{eqnarray}\label{phiaphibphic}
|\phi_a\rangle&=&|H\rangle_{a_p}\otimes|V\rangle_{a_p}\xrightarrow{VBS_1}(\sqrt{T}|H\rangle_{a_{p1}}+\sqrt{1-T}|H\rangle_{a_{p2}})\otimes(\sqrt{T}|V\rangle_{a_{p1}}+\sqrt{1-T}|V\rangle_{a_{p2}}),\nonumber\\
|\phi_b\rangle&=&|H\rangle_{b_p}\otimes|V\rangle_{b_p}\xrightarrow{VBS_2}(\sqrt{T}|H\rangle_{b_{p1}}+\sqrt{1-T}|H\rangle_{b_{p2}})\otimes(\sqrt{T}|V\rangle_{b_{p1}}+\sqrt{1-T}|V\rangle_{b_{p2}}),\nonumber\\
|\phi_c\rangle&=&|H\rangle_{c_p}\otimes|V\rangle_{c_p}\xrightarrow{VBS_3}(\sqrt{T}|H\rangle_{c_{p1}}+\sqrt{1-T}|H\rangle_{c_{p2}})\otimes(\sqrt{T}|V\rangle_{c_{p1}}+\sqrt{1-T}|V\rangle_{a_{p2}}).
\end{eqnarray}
The transmitted photons pass through the lossy channels with the transmission efficiency of $\eta_t$ and are stored in $QM_{A2}$, $QM_{B2}$, and $QM_{C2}$ at the center measurement station, respectively. The reflected photons are directed into $QM_{A1}$, $QM_{B1}$, and $QM_{C1}$, respectively. Each QM continuously monitors the arrival of single photon and replaces stored photon with newer arriving one. At this point, the whole quantum state in the six QMs $|\Phi\rangle=|\phi_a\rangle\otimes|\phi_b\rangle\otimes|\phi_c\rangle$ collapses to
\begin{eqnarray}\label{collapsePhi1}
|\Phi\rangle&\rightarrow&\eta_M\sqrt{\eta_tT(1-T)}\left(|H\rangle|V\rangle+|V\rangle|H\rangle\right)_{QM_{A1}QM_{A2}}\otimes\eta_M\sqrt{\eta_tT(1-T)}\left(|H\rangle|V\rangle+|V\rangle|H\rangle\right)_{QM_{B1}QM_{B2}}\nonumber\\
&\otimes&\eta_M\sqrt{\eta_tT(1-T)}\left(|H\rangle|V\rangle+|V\rangle|H\rangle\right)_{QM_{C1}QM_{C2}},
\end{eqnarray}
where $\eta_M$ denotes the storage efficiency of the quantum memory, and the quantum memory structure is shown in Fig. 1 (c) in the main text.

\emph{\textbf{Step 3}}. When all the QMs are loaded with single photons, the three users send the photons which are stored in the $QM_{A2}$, $QM_{B2}$, and $QM_{C2}$ through the paths $a_{p1}'$, $b_{p1}'$ and $c_{p1}'$ to the GSM module at the same time. After the entanglement swapping, the whole quantum state becomes
\begin{eqnarray}\label{collapsePhi2}
|\Phi\rangle&\rightarrow&\left(\eta_M\sqrt{\eta_tT(1-T)}\right)^3\times\Big(|GHZ_1^+\rangle_{A_{in}B_{in}C_{in}}|GHZ_1^+\rangle_{QM_{A1}QM_{B1}QM_{C1}}\nonumber\\
&-&|GHZ_1^-\rangle_{A_{in}B_{in}C_{in}}|GHZ_1^-\rangle_{QM_{A1}QM_{B1}QM_{C1}}+|GHZ_2^+\rangle_{A_{in}B_{in}C_{in}}|GHZ_2^+\rangle_{QM_{A1}QM_{B1}QM_{C1}}\nonumber\\
&-&|GHZ_2^-\rangle_{A_{in}B_{in}C_{in}}|GHZ_2^-\rangle_{QM_{A1}QM_{B1}QM_{C1}}+|GHZ_3^+\rangle_{A_{in}B_{in}C_{in}}|GHZ_3^+\rangle_{QM_{A1}QM_{B1}QM_{C1}}\nonumber\\
&-&|GHZ_3^-\rangle_{A_{in}B_{in}C_{in}}|GHZ_3^-\rangle_{QM_{A1}QM_{B1}QM_{C1}}+|GHZ_4^+\rangle_{A_{in}B_{in}C_{in}}|GHZ_4^+\rangle_{QM_{A1}QM_{B1}QM_{C1}}\nonumber\\
&-&|GHZ_4^-\rangle_{A_{in}B_{in}C_{in}}|GHZ_4^-\rangle_{QM_{A1}QM_{B1}QM_{C1}}\Big),
\end{eqnarray}
where the eight three-photon polarization GHZ states are given by
\begin{eqnarray}\label{8GHZ}
|GHZ_1^\pm\rangle&=&\frac{1}{\sqrt{2}}\left(|HHH\rangle\pm |VVV\rangle\right),\quad\quad|GHZ_2^\pm\rangle=\frac{1}{\sqrt{2}}\left(|HHV\rangle\pm |VVH\rangle\right),\nonumber\\
|GHZ_3^\pm\rangle&=&\frac{1}{\sqrt{2}}\left(|HVH\rangle\pm |VHV\rangle\right),\quad\quad|GHZ_4^\pm\rangle=\frac{1}{\sqrt{2}}\left(|HVV\rangle\pm |VHH\rangle\right).
\end{eqnarray}

The linear-optical GSM module can only distinguish two GHZ states $|GHZ_1^\pm\rangle$. Only when the GHZ state measurement is successful, the photons stored in $QM_{A1}$, $QM_{B1}$, and $QM_{C1}$ can successfully establish the three-partite entanglement, which is called the successful entanglement event. When the GHZ state measurement result is $|GHZ_1^-\rangle$, a phase-flip operation must be applied before Alice's measurement device to convert $|GHZ_1^-\rangle_{QM_{A1}QM_{B1}QM_{C1}}$ into $|GHZ_1^+\rangle_{QM_{A1}QM_{B1}QM_{C1}}$. In this way, the three users can deterministically construct the three-partite entanglement channels in $|GHZ_1^+\rangle_{QM_{A1}QM_{B1}QM_{C1}}$.

The above three steps can eliminate the effect of photon transmission loss on the quality of the three-partite entanglement channels.

\section{The practical key generation efficiency of QMA SPS DI QSS}
\subsection{Fully loaded efficiency $E_m$}
We adopt the all-optical polarization-insensitive storage loops QM \cite{QM1} with the structure of Fig. 1 (c) in main text. Here, we configure that each QM can store a single photon for at most $N$ photon pulse intervals. We analyze the efficiency $E_m$ that all the six QMs are loaded by single photons per unit time.

In Step 3, each user needs to send one of the two photons to the QM at the center measurement station through the lossy channel, while stores the other  reflected photon in the local QM. The probability $P_s$ that a user achieves such a successful two-photon separation event is denoted as
\begin{eqnarray}\label{Ps}
P_s=2\eta_tT(1-T),
\end{eqnarray}
where the transmission efficiency of the lossy channel $\eta_t=10^{-\alpha d/10}$ ($d$ denotes the photon transmission distance, and $\alpha=0.2$ dB/km for the standard optical fiber). It is easy to find that $P_s$ reduces with the growth of communication distance $d$.

We assume that three users achieve successful separation events at the $(l+1)$th, $(m+1)$th, and $(n+1)$th pulse interval, respectively, where $l,m\leq n$ and the actual storage pulse interval $n\leq N$. Therefore, the probability that all six QMs are fully loaded with photons at the $(n+1)$th photon pulse interval is
\begin{eqnarray}\label{Pn+1}
P(n+1)&=&\left(\sum_{m=0}^{n} \left(\sum_{l=0}^{n} P_{s}\left(1-P_{s}\right)^{n-l}\left(\eta_{M}^{2}\right)^{n-l}\right) P_{s}\left(1-P_{s}\right)^{n-m}\left(\eta_{M}^{2}\right)^{n-m}\right) P_{s}\left(1-P_{s}\right)^{n}\nonumber\\
&=&P_{s}^{3}\left[\eta_{M}^{2}\left(1-P_{s}\right)\right]^{2 n}\left(1-P_{s}\right)^{n} \sum_{l=0}^{n} \sum_{m=0}^{n}\left(\frac{1}{\eta_{M}^{2}\left(1-P_{s}\right)}\right)^{l+m},
\end{eqnarray}
where $\eta_M$ denotes the storage efficiency of each QM. Given the complexity of performing multiple summation and optimization, we simplify Eq.~\eqref{Pn+1} by decomposing it into the summation of two geometric progressions, where both of their common ratios are $\frac{1}{\eta_M^2(1-P_s)}\neq 1$. Thus, Eq.~\eqref{Pn+1} can be simplified as
\begin{eqnarray}\label{Pn+12}
P(n+1)&=&P_s^3 \left[\eta_M^2 (1-P_s)\right]^{2n} (1-P_s)^n \sum_{m=0}^n \frac{\left( \frac{1}{\eta_M^2 (1-P_s)} \right)^m \left[ 1 - \left( \frac{1}{\eta_M^2 (1-P_s)} \right)^{n+1} \right]}{1-\frac{1}{\eta_M^2 (1-P_s)}} \nonumber\\
&=& P_s^3 [\eta_M^2 (1-P_s)]^{2n} (1-P_s)^n \left[ \frac{1 - \left( \frac{1}{\eta_M^2 (1-P_s)} \right)^{n+1}}{1-\frac{1}{\eta_M^2 (1-P_s)}} \right]^2 = P_s^3 (1-P_s)^n \left[ \frac{(\eta_M^2 (1-P_s))^n - 1}{1 - \frac{1}{\eta_M^2 (1-P_s)}} + 1 \right]^2.
\end{eqnarray}
Therefore, for all cases with $n\leq N$, the total probability that all QMs are fully loaded with photons is
\begin{eqnarray}\label{PtN+1}
P_t(N+1)=\sum_{n=0}^{N}P(n+1).
\end{eqnarray}

\begin{figure}
\includegraphics[scale=0.34]{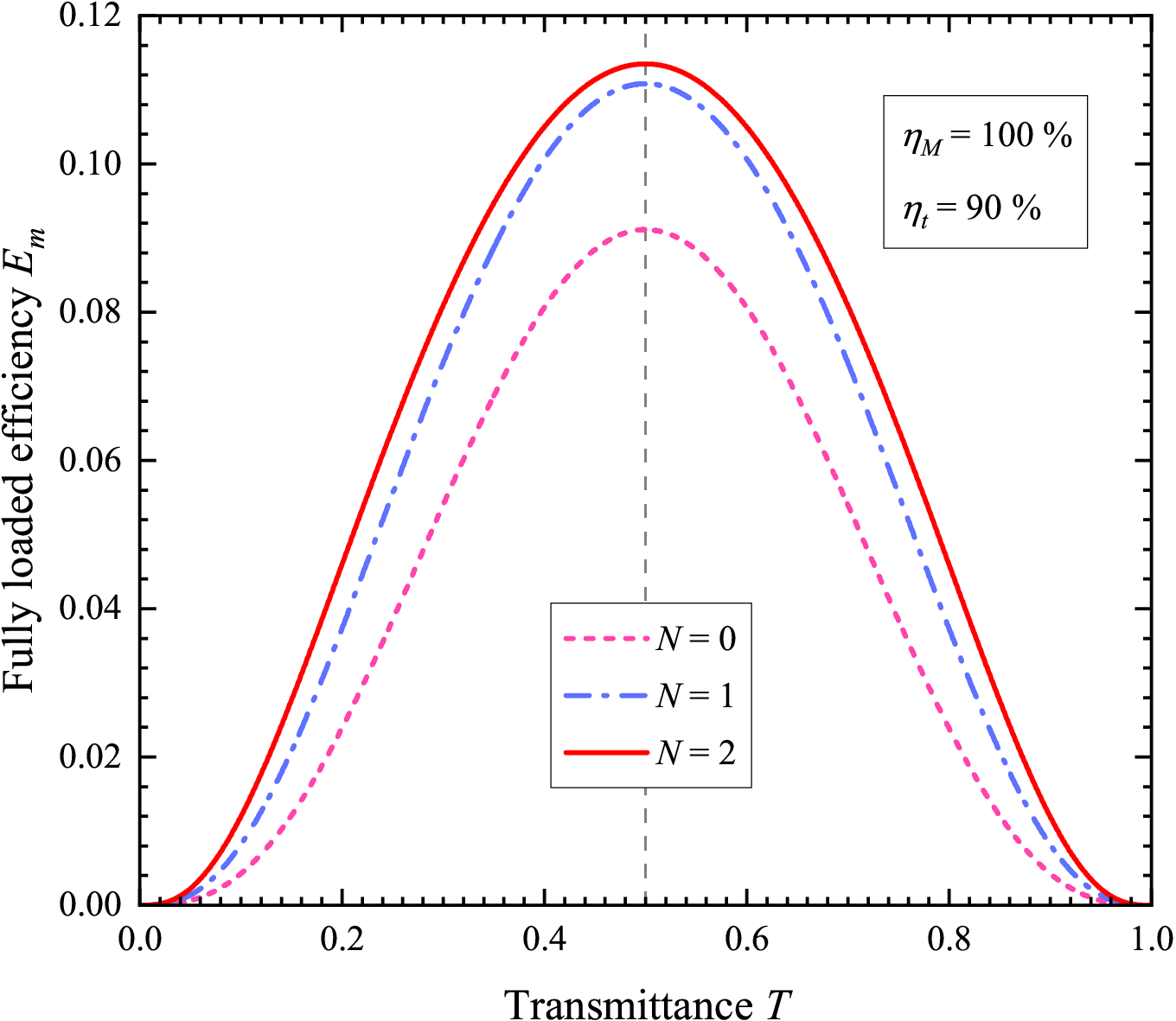}
\caption{The fully loaded efficiency $E_m$ as a function of transmittance $T$, where  we assume the storage efficiency $\eta_M=100\%$ and the photon transmission efficiency $\eta_t=90\%$.}
\label{figS1}
\end{figure}

Since, the number of consumed photon pulses is proportional to the actual storage pulse interval $n$, we need to consider the average photon pulse consumption per unit time, denoted as $P_w$, which represents the statistical average over all $n$. For simplicity, we divide the analysis into two cases: $n<N$ and $n=N$. In the first case, the three users complete a QMs fully loaded event in advance, and each single-photon source must emit $n+1$ photon pulses. Therefore, the  photon pulse interval consumption is expressed as
\begin{eqnarray}\label{PwN}
P_w(\Sigma N)=\sum_{n=0}^{N-1}(n+1)P(n+1).
\end{eqnarray}
In the second case, regardless of whether all QMs are successfully loaded with single photons, each SPS must emit $N+1$ photon pulses. Therefore, the photon pulse interval consumption is expressed as
\begin{eqnarray}\label{PwN+1}
P_w(N+1)=(N+1)(1-P_t(N)).
\end{eqnarray}
Finally, we obtain the total photon pulse interval consumption as $P_w=P_w (\Sigma N)+P_w (N+1)$. The repetition frequency of the single-photon source is $R_{rep}$. Therefore, the three users can attempt approximately $R_{rep}/P_w$ rounds per unit time. Since the success probability of a fully loaded event is $P_t(N+1)$, the expected number of successful events per unit time is about $P_t(N+1)R_{rep}/P_w$. We can obtain the full loaded efficiency $E_m$  as
\begin{eqnarray}\label{Em}
E_m=\frac{P_t(N+1)R_{rep}}{P_w R_{rep}}=\frac{P_t(N+1)}{P_w(\Sigma N)+P_w(N+1)}.
\end{eqnarray}

To maximize the efficiency $E_m$, Fig.~\ref{figS1} provides $E_m$ as a function of the transmittance $T$. We assume the storage efficiency $\eta_M=100\%$ and the photon transmission efficiency $\eta_t=90\%$. It can be found that $E_m$ reaches its maximum value when $T=0.5$, indicating that $T=0.5$ is the optimal value of the transmittance $T$.

\subsection{Lower bound on the total secure key rate $R_\infty$}
In our QMA SPS DI QSS protocol, the practical key generation efficiency $E_c$ can be calculated by
\begin{eqnarray}\label{Ec1}
E_c=(1-P_c)P_{GHZ}R_{rep}E_mR_\infty,
\end{eqnarray}
where $P_c=50\%$, $P_{GHZ}=1/4$, and $R_{rep}=10$ MHz \cite{source1}. The total secure key rate $R_\infty$ refers to the ratio of the extractable key length to the number of key generation rounds.

This subsection briefly computes the total secure key rate $R_\infty$. Alice (Charlie) chooses $A_1$ and $A_2$ ($C_1$ and $C_2$) measurement basis with probabilities $p$ and $\bar{p}=1-p$, respectively. We set $p=50\%$. In the asymptotic limit of a large number of rounds, the total secure key rate $R_\infty$ is given by
\begin{eqnarray}\label{rinfty1}
R_\infty=p^2 r_{111}=p^2\left[H\left(A_1|E\right)-H\left(A_1|B_1,C_1\right)\right],
\end{eqnarray}
where $H(|)$ is the von Neumann conditional entropy. $H(A_1|E)$ quantifies Eve's uncertainty about Alice's key. $H(A_1|B_1,C_1)$ quantifies the uncertainty about Bob's and Charlie's keys with respect to the measurement basis combination $\{A_1B_1C_1\}$. Reference \cite{DIQSS3} computes the lower bound of DI QSS's $R_\infty$ under collective attack (\emph{i.e.}, obeying the independent and identical distribution assumption) by using $\{A_1B_1C_1\}$ measurement basis combination as
\begin{eqnarray}\label{rinfty2}
R_\infty=p^2 r_{111}\geq p^2\left[g\left(\sqrt{S^2/4-1}\right)-h(\delta)\right],
\end{eqnarray}
where we define the function $g(x)=1-h(\frac{1}{2}+\frac{1}{2}x)$, the binary Shannon entropy $h(x)=-x\log_{2}{x}-(1-x)\log_{2}{(1-x)}$, $S$ is the CHSH polynomial values between Alice's and Bob's measurement results, and $\delta$ is the total quantum bit error rate (QBER).

\begin{figure}
\includegraphics[scale=0.34]{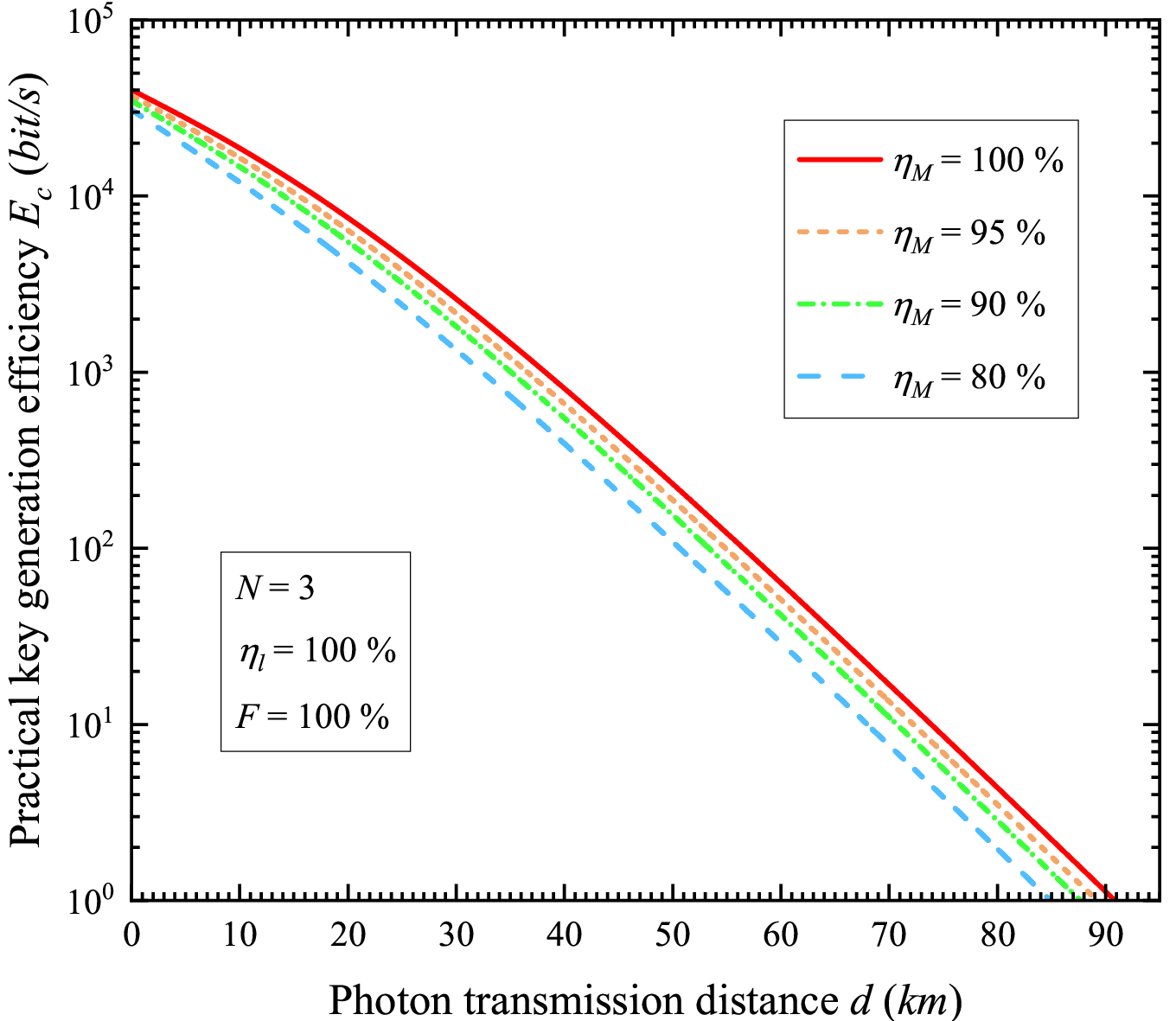}
\caption{The practical key generation efficiency $E_c$ as a function of the storage efficiency $\eta_M$. We set the maximum storage pulse interval $N=3$, the fidelity $F=100\%$ and the local efficiency $\eta_l=100\%$.}
\label{figS2}
\end{figure}

The entangled photons suffer from noise and occur photon local loss and decoherence. Here, we consider the white noise model. During local measurement, photons may be lost with probability $\bar{\eta_l}$ and change to the eight noisy GHZ states in Eq.~\eqref{8GHZ} with equal probability $\frac{1-F}{8}$. Therefore, the theoretical value of the CHSH polynomial between Alice's and Bob's measurement results is given by
\begin{eqnarray}\label{Swhite}
S=2\sqrt{2}F\eta_l^3,
\end{eqnarray}
where the fidelity $F$ is the probability that no error occurs in the photonic state and $\eta_l$ is the local efficiency. The total QBER $\delta$ obtained from the white noise model combining with the photon local loss \cite{DIQSS3} can be calculated as
\begin{eqnarray}\label{totaldelta}
\delta=\frac{1-F}{2}\eta_l^3+1-\eta_l^{3}=1-\frac{1}{2}\eta_l^3-\frac{1}{2}\eta_l^3F.
\end{eqnarray}

In conclusion, we substitute Eqs.~\eqref{Em}-\eqref{totaldelta} into Eq.~\eqref{Ec1} to obtain a lower bound of the practical key generation efficiency as
\begin{eqnarray}\label{EC2}
E_c&=&(1-P_c)P_{GHZ}R_{rep}E_mR_\infty \geq \frac{1}{8}R_{rep}E_mp^2\left[g\left(\sqrt{2F^2\eta_l^6-1}\right)-h\left(1-\frac{1}{2}\eta_l^3-\frac{1}{2}\eta_l^3F\right)\right].
\end{eqnarray}
Thereby, $E_c$ decreases as the fidelity $F$ decreases \cite{DIQSS3,DIQSS4}. To obtain a positive practical key generation efficiency, the critical value of $F$ is about $81.54\%$ when $\eta_l=100\%$.

Since the QMA SPS DI QSS is similar to the multiparty quantum repeater structure, the storage efficiency $\eta_M$ is closely related to the practical key generation efficiency. However, QMA SPS DI QSS imposes lower requirements on the QM module. Specifically, it employs a polarization-insensitive storage loop \cite{QM1}, which significantly reduces the cost of the experimental demonstration. In Fig.~\ref{figS2}, it can be found that as the storage efficiency $\eta_M$ decreases, the practical key generation efficiency $E_c$ of QMA SPS DI QSS decreases at a fixed photon transmission distance.

\section{Relaxing the performance requirement for local devices}

In this subsection, we adopt the advanced random key generation basis strategy to relax the performance requirements on local devices in QMA SPS DI QSS protocol.

We first consider the postselection strategy, in Step 4, we randomly binary the detector non-click result $\perp$ to $+1$ or $-1$ \cite{DIQSS3}. Thereby, the total QBER in Eq. \eqref{totaldelta} becomes
\begin{eqnarray}\label{totaldeltap}
\delta_p=\frac{1-F}{2}\eta_l^3+\frac{1}{2}(1-\eta_l^{3})=\frac{1}{2}-\frac{1}{2}\eta_l^3F.
\end{eqnarray}

\begin{figure}
\includegraphics[scale=0.34]{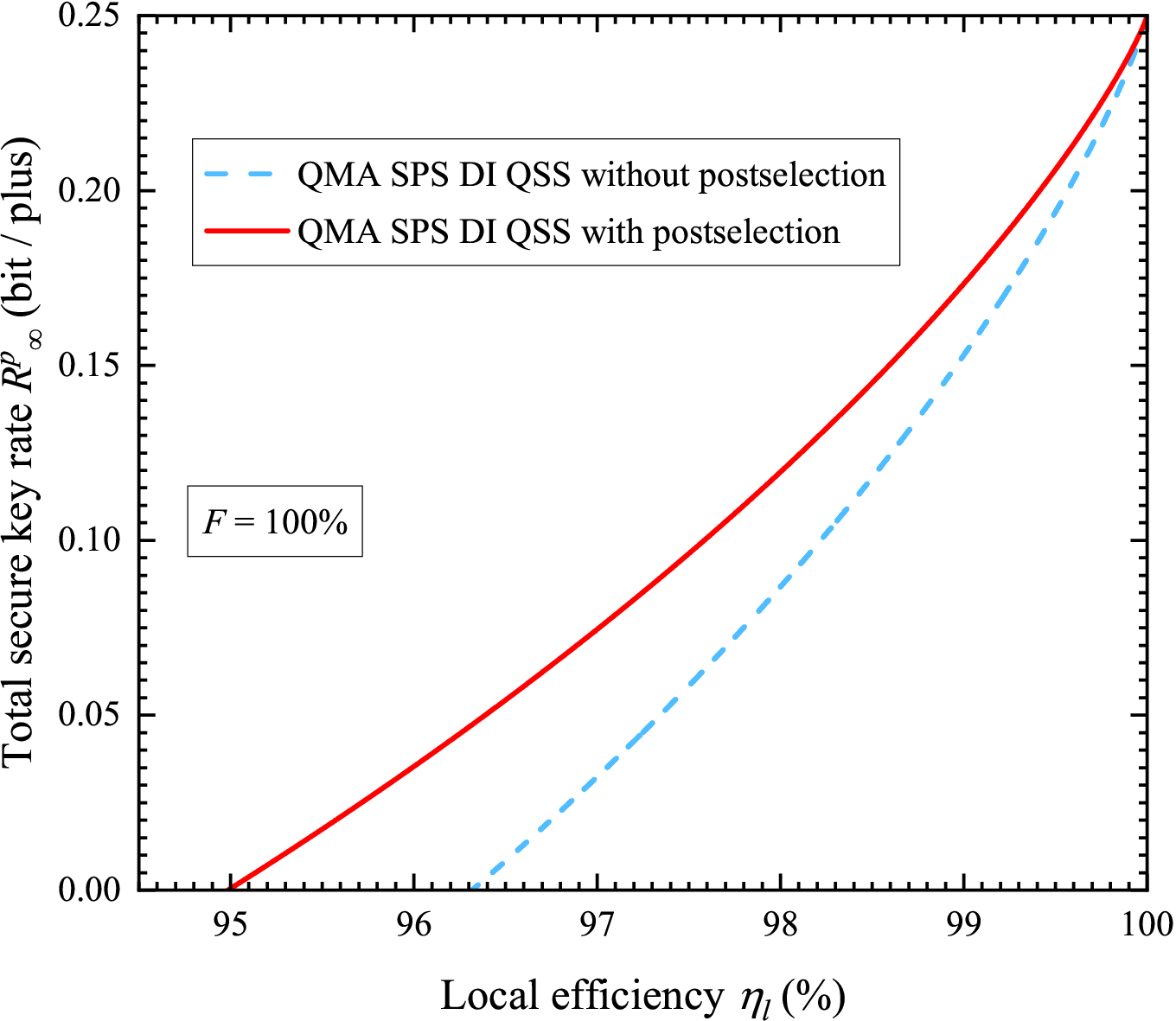}
\caption{The secure key rate of QMA SPS DI QSS as a function of local efficiency with and without the postselection strategy, where we assume the fidelity $F=100\%$.}
\label{figS3}
\end{figure}

The postselection strategy does not affect the CHSH value $S=2\sqrt{2}F\eta_l^3$ between Alice's and Bob's measurement results. By substituting $\delta_p$ into Eq.~\eqref{rinfty2}, we can obtain the total secure key rate $R_\infty^p$ after adopting the postselection strategy. In Fig.~\ref{figS3}, we demonstrate the effect of the postselection strategy on the total secure key rate $R_\infty^p$ with respect to the local efficiency $\eta_l$. We can estimate the local efficiency threshold $\eta_l$ of QMA SPS DI QSS is reduced from 96.32\% to 94.99\% after adopting the postselection strategy, where the GHZ state fidelity $F=100\%$. Therefore, the postselection strategy effectively reduces the performance requirements of the local devices.

Then, we adopt the advanced random key generation basis strategy \cite{DIQSS4}, which integrates three techniques including noise preprocessing, postselection, and random key basis strategies. In the key generation stage of Step 4, instead of only using a fixed measurement basis combination $\{A_1B_1C_1\}$ to generate key bits, three users randomly select two measurement basis combinations $\{A_1B_1C_1\}$ and $\{A_2B_1C_2\}$ to generate key bits, which is called the random key generation basis strategy. Furthermore, for Alice's measurement result, she may flip it with probability $q$ ($+1\leftrightarrow-1$) and announce this flip probability in the error correction stage, which is called the noise preprocessing strategy. These two active improvement strategies effectively increase Eve's total uncertainty $H(A|E)$ about the key generation basis and measurement results, thus increasing the total secure key rate $R_\infty$.

After adopting the advanced random key generation basis strategy, the total secure key rate is changed from Eq.~\eqref{rinfty1} to
\begin{eqnarray}\label{rinftyar1}
R^{ar}_\infty=p^2 r_{111}+\bar{p}^2 r_{212},
\end{eqnarray}
where $r_{111}=H(A_1|E)_{ar}-H(A_1|B_1,C_1)_{ar}$ and $r_{212}=H(A_2|E)_{ar}-H(A_2|B_1,C_2)_{ar}$. Rearranging the above equation, we obtain
\begin{eqnarray}\label{rinftyar2}
R^{ar}_\infty=(p^2+\bar{p}^2)\left[H(A|E)_{ar}-\left(\lambda H(A_1|B_1,C_1)_{ar}+\bar{\lambda}H(A_2|B_1,C_2)_{ar}\right)\right],
\end{eqnarray}
where the total key secrecy rate of Eve is $H(A|E)_{ar}=\lambda H(A_1|E)_{ar}+\bar{\lambda}H(A_2|E)_{ar}$. The matching weight of the first key generation basis $\{A_1B_1C_1\}$ is $\lambda=p^2/(p^2+\bar{p}^2)$, and the second key generation basis $\{A_2B_1C_2\}$ is $\bar{\lambda}=1-\lambda$.

Ref. \cite{DIQSS4} numerically estimates the lower bound of the total key secrecy rate $H(A|E)_{qp}$ of Eve with the advanced random key generation basis strategy as
\begin{eqnarray}\label{keysecrecyrate}
H\left(A|E\right)_{ar}\ge g\left(\tilde{E}_\lambda(S),q\right),
\end{eqnarray}
where $S=2\sqrt{2}F\eta_l^3$ and $\tilde{E}_\lambda(S)\equiv \sqrt{\tilde{E}_\lambda(S)^2}$ can be obtained by solving the optimization problem
\begin{eqnarray}\label{maxproblem}
E_\lambda(S)^2=\max& s^2g^2+c^2h^2+2(2\lambda-1)scgh\varDelta, \nonumber\\
s.t.& cg+sh\geq S/2, \nonumber\\
& g^2 \leq 1, \nonumber\\
& h^2 \leq 1, \nonumber\\
& (1-g^2)(1-h^2)\geq g^2h^2\varDelta^2, \nonumber\\
& c^2+s^2=1, \nonumber\\
& \varDelta ^{2} \leq 1,
\end{eqnarray}
with the noise preprocessing entropy function $g(x,q)$ as
\begin{eqnarray}\label{gxq}
g(x,q)=1-h\left(\frac{\sqrt{x^2/4-1}}{2}+\frac{1}{2}\right)+h\left[\frac{\sqrt{\left(1-2q\right)^2+4q\left(1-q\right)\left(x^2/4-1\right)}}{2}+\frac{1}{2}\right].
\end{eqnarray}

\begin{figure}
\includegraphics[scale=0.34]{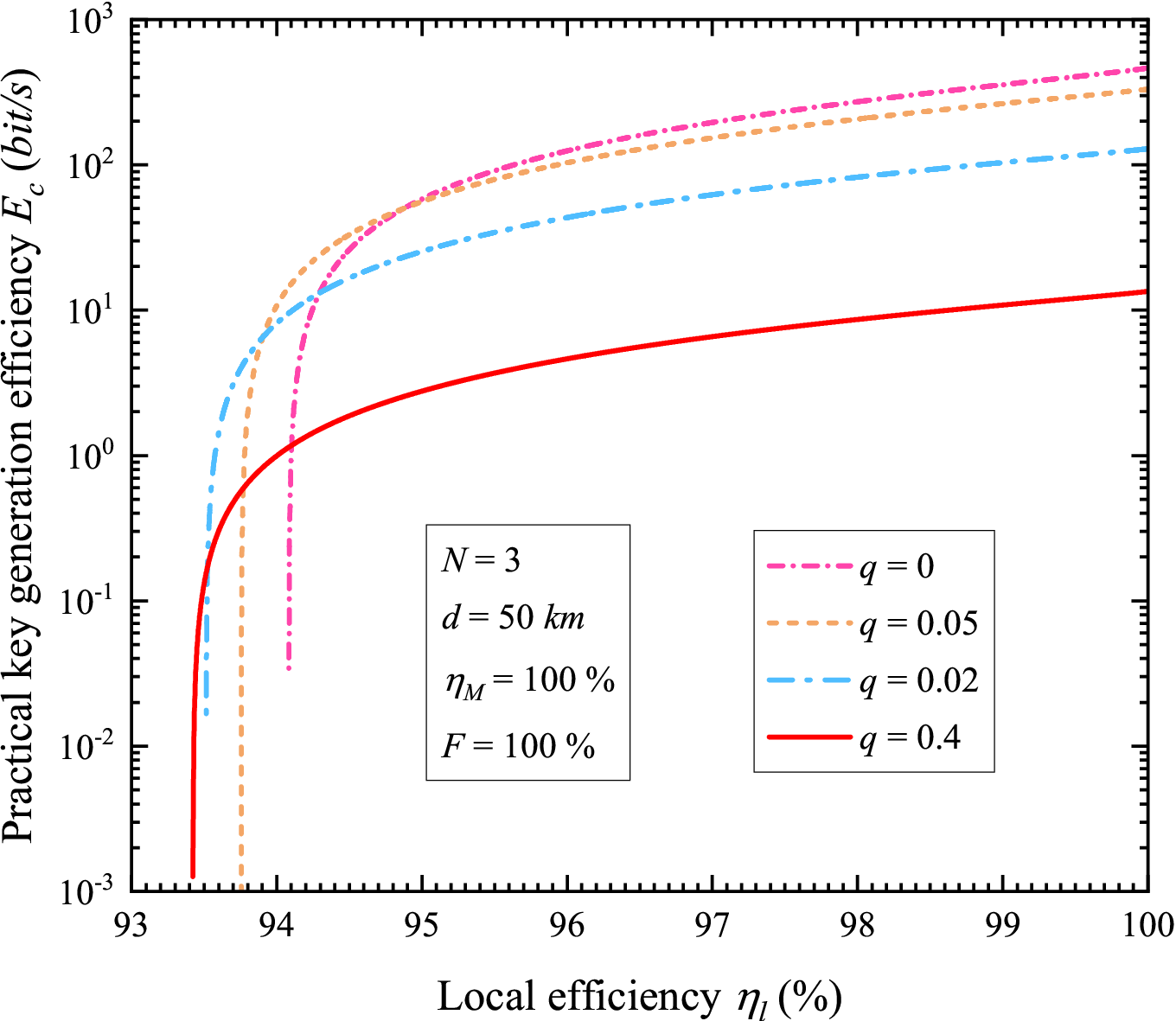}
\caption{The practical key generation efficiency $E_c$ as a function of the local efficiency $\eta_l$ for the QMA SPS DI QSS protocol with the advanced random key generation basis strategy ($q=0,0.05,0.2,0.4$), where we set the photon transmission distance $d=50$ km, the storage efficiency $\eta_M=100\%$ and the fidelity $F=100\%$.}
\label{figS4}
\end{figure}

With the advanced random key generation basis, the total QBER $\delta_{ar}$ is given by
\begin{eqnarray}\label{totalQBERar}
\delta_{ar}=q+(1-2q)\delta_p=q+(1-2q)(\frac{1}{2}-\frac{1}{2}\eta^3_lF).
\end{eqnarray}
As a result, the irrelevance of the reconstructed key after Bob and Charlie cooperate with the shared key of the Alice is $H(A_1|B_1,C_1)_{ar}=H(A_2|B_1,C_2)_{ar}=h(\delta_{ar})$.

According to Eq.~\eqref{rinftyar2}, the lower bound of the secure key rate $R_\infty^{ar}$ after adopting the advanced random key generation basis strategy is
\begin{eqnarray}\label{rinftyar}
R^{ar}_\infty \ge \left(p^2+\bar{p}^2\right)\left[g\left(\tilde{E}_\lambda\left(S\right),q\right)- h(\delta_{ar})\right].
\end{eqnarray}

Substituting Eqs.~\eqref{keysecrecyrate} and \eqref{totalQBERar} into Eq.~\eqref{rinftyar}, the practical key generation efficiency $E_c^{ar}$ of QMA SPS DI QSS with advanced random key generation basis strategy is
\begin{eqnarray}\label{ECar}
E_c^{ar}&=&(1-P_c)P_{GHZ}R_{rep}E_mR_\infty^{ar}\nonumber\\
&\geq & \frac{1}{8}R_{rep}E_m\left(p^2+\bar{p}^2\right)\left[g\left(\tilde{E}_\lambda\left(2\sqrt{2}F\eta_l^3\right),q\right)-h\left(q+(1-2q)\left(\frac{1}{2}-\frac{1}{2}\eta^3_lF\right)\right)\right].
\end{eqnarray}

In Fig.~\ref{figS4}, we show the impact of the advanced random key generation basis strategy on the local detection efficiency threshold of QMA SPS DI QSS when the distance between each user and the central station is $d=50$ km. A clear trade-off is observed between the level $q$ of noise preprocessing and the practical key generation efficiency. As $q$ increases, the advanced random key generation basis strategy effectively lowers the required local detection efficiency threshold $\eta_l$. However, this gain comes at the cost of a reduced practical key generation efficiency. When the noise preprocessing level $q\rightarrow50\%$, the local efficiency threshold $\eta_l$ decreases to 93.41\%, demonstrating a significant improvement in robustness against local loss.